\documentclass[aps,nofootinbib,amssymb,amsmath,
twocolumn,
showpacs,showkeys]{revtex4}
\usepackage{graphicx}
\usepackage{epsfig}
\input{psfig.sty}
\newcommand{\bastar}{\begin{eqnarray*}}
\newcommand{\eastar}{\end{eqnarray*}}
\newskip\humongous \humongous=0pt plus 1000pt minus 1000pt

\newif\ifdtup

\relax
\newcommand{\be}{\begin{equation}}
\newcommand{\ee}{\end{equation}}
\newcommand{\ba}{\begin{array}}
\newcommand{\ea}{\end{array}}
\newcommand{\bea}{\begin{eqnarray}}
\newcommand{\eea}{\end{eqnarray}}
\newcommand{\pro}{\partial}
\newcommand{\oneg}{\difrac{1}{g}}

\newcommand{\difrac}{\displaystyle\frac}
\newcommand{\nn}{\nonumber}

\newcommand{\circe}{{\stackrel{\circ}{e}}}
\newcommand{\circphi}{{\stackrel{\circ}{\varphi}}}
\newcommand{\circR}{{\stackrel{\circ}{R}}}
\newcommand{\tomega}{\tilde \omega}
\newcommand{\abc}{\alpha \beta \gamma}

\newcommand{\n}{\vec n}

\newcommand{\A}{{\vec A}}
\newcommand{\B}{{\vec B}}

\newcommand{\F}{{\vec F}}

\newcommand{\vOm}{{\bf \Omega}}

\newcommand{\hn}{{\hat n}}

\begin{document}
\title{Vacuum tunneling in gravity}
\bigskip
\author{Y.M. Cho}
\email{ymcho@unist.ac.kr}
\affiliation{School of Physics, College of Natural Sciences, \\
Seoul National University, Seoul, 151-742, Korea}
\affiliation{ School of Electrical and Computer Engineering\\
    Ulsan National Institute of Science and Technology,
    Ulsan 689-798, Korea}
\author{D.G. Pak}
\email{dmipak@gmail.com}
\affiliation{Center for Theoretical Physics, Seoul National
University, Seoul 151-742, Korea}
\affiliation{Institute of Applied Physics, Uzbekistan National University,
Tashkent 100174, Uzbekistan}
\begin{abstract}
Topologically non-trivial vacuum structure in gravity models with
Cartan variables (vielbein and contortion) is considered.
We study the possibility of vacuum space-time tunneling in Einstein gravity
assuming that the vielbein may play a fundamental role in quantum
gravitational phenomena. It has been shown that in the case of $RP^3$
space topology the tunneling between non-trivial topological vacuums
can be realized by means of Eguchi-Hanson gravitational instanton.
In Riemann-Cartan geometric approach to quantum gravity the vacuum
tunneling can be provided by means of contortion quantum fluctuations.
We define double self-duality condition for the contortion and
give explicit self-dual configurations which can contribute
to vacuum tunneling amplitude.
\end{abstract}

\maketitle

{\bf I. Introduction}

The problem of quantum gravity remains a great unresolved puzzle in theoretical physics.
There are several very sophisticated models based on superstrings, loop quantum gravity
\cite{rovelli, smolin}, Euclidean gravity (see recent book \cite{hamber} and references therein)
and others which reveal new perspectives in last decades towards more deep understanding
of the nature of quantum gravitation. In any quantum field theory the problem of the
vacuum is one of most important issues which must be studied
first to make firm foundation of the theory.
Recently the classification of non-trivial
topological vacuums in Einstein gravity has been proposed \cite{cho07}.
The topological vacuum structure and the
possibility of vacuum tunneling in Euclidean gravity were considered
in late 70s \cite{hawking,gibb-hawk} with a concluding note
that the topological vacuums are separated
from each other by an infinite energy barrier which
makes the tunneling impossible.
This conclusion is valid only under certain assumptions
about the global topological properties of the
gravitational vacuum.
It is known that the Einstein gravity can be formulated either
in terms of the metric tensor or in tetrad (vielbein) formalism.
In the absence of fermions both formulations seemed to be equivalent.
However, even in a pure gravity without matter the vielbein may play a more fundamental role
than the metric. Notice, that in the geometric theory of defects the vielbein
represents the basic independent variable, especially in presence of dislocations
\cite{moraes,katanaev}. One should mention as well, that the idea of relating dislocations
to the torsion tensor had appeared in 1950s \cite{kondo}. This provides interesting links to
possibility that quantum gravity could be related to torsion within Riemann-Cartan geometry.
In the present paper we will demonstrate that vielbein provides
a non-trivial topological vacuum structure which can be manifested
through the quantum tunneling effect. We study also simple torsion instanton
configurations which can provide similar quantum tunneling effects.

We use the gauge formalism based on local Lorentz symmetry
as a fundamental symmetry of quantum gravity
 \cite{uti,kib, sciama,carmelli,uti-fuku,prd76a,prd76b}.
In Section II we consider the vacuum concept in Einstein gravity
and explore the non-trivial topological vacuum structure
with a simple ansatz for the vielbein.
In Section III we establish connection between topologically non-equivalent
classes of $SU(2)$ gauge potentials and corresponding non-equivalent
classes of vielbein in Einstein gravity.
We show explicitly that
the Eguchi-Hanson gravitational instanton \cite{E-H}
can provide vacuum tunneling in Einstein gravity
in a case of base space topology $RP^3$.
Section IV is devoted to the vacuum structure and
self-dual contortion configurations in
gravity models within Riemann-Cartan geometric formalism.
The last Section V contains discussion on possible
physical implications of the non-trivial topological vacuum
structure in Einstein gravity.

{\bf II. Vacuum in Einstein gravity}

In Lorentz gauge approach to generalized theory of gravity the basic independent
variables are represented by vielbein $e_{am}$ and
Lorentz gauge connection $\gamma_{mcd}$ (i,k,l,... are used for world space-time indices
and a,b,c,... for Lorentz group indices).
The topologically non-equivalent classes
of vielbein and gauge connection are classified
by the non-trivial homotopy group $\pi_3 (SO(1,3))=\pi_3 (SO(3))$
in a classical Lorentz gauge gravity, and by
$\pi_3(SO(4))=\pi_3 (SU(2)\times SU(2))$ in Euclidean
formulation of the gravity which is relevant to description of
quantum fluctuations.
So that the topological structure
is determined by configurations of both fields, vielbein and gauge
connection. In Riemannian geometry the Lorentz gauge
connection is not an independent geometrical object,
it is defined by Levi-Civita connection constructed in terms
of vielbein. In that case the space-time geometry is completely determined
by the vielbein. In Riemann-Cartan geometry the contortion alone
(as an independent part of the gauge Lorentz connection)
alone can provide a non-trivial topological structure in the theory,
even in a case of the flat space-time.

Despite on lack of renormalizability of Einstein
gravity there is still a possibility that a
consistent quantum theory of gravity might
exist in non-perturbative regime.
In our analysis of the vacuum tunneling problem
we will use an assumption that the vielbein is
a more fundamental field than the metric,
and it can represent dynamic degrees of freedom of
quantum gravity. We will concentrate mainly on vacuum
space-time structure caused by vielbein configuration space.

Let us start with main outlines
of the general structure of Riemann-Cartan
geometry.
The Lorentz gauge connection $\gamma_{mcd}$
can be decomposed into Levi-Civita spin connection $\varphi_{mcd}(e)$
and contortion $K_{mcd}$
\bea
\gamma_{mcd}=\varphi_{mcd}(e)+K_{mcd}. \label{split}
\eea
The Levi-Civita connection is defined in terms of vielbein as follows
\bea
\varphi_{m a}^{~~b}(e)&=&\difrac{1}{2} (
e_a^n \pro_m e_n^b
-e^{n b} e_m^c \pro_a e_{n c}+\nn \\
&&\pro^b e_{m a}-(a\leftrightarrow b)).\label{LeviCivita}
\eea
The vielbein $e_m^a$ forms the basis of differential 1-forms
in cotangent bundle with the base space-time manifold
$M^4$ and with the structure Lorentz group.
The metric of the space-time manifold is determined
through the relationship
\bea
g_{mn}=e_m^a e_{na}. \label{def:metric}
\eea

Covariant derivatives acting on Lorentz and world vectors
are defined with the help of Lorentz spin connection and
Riemann-Cartan connection $\Gamma_{nm}^k$ respectively
\bea
D_m V_a & =& \pro_m V_a+ \gamma_{ma}^{~~b} V_b, \nn \\
D_m V_n &=& \pro_m V_n - \Gamma_{nm}^k V_k.
\eea
The Riemann-Cartan connection $\Gamma_{nm}^k$
can be decomposed into the Christoffel symbol $\hat \Gamma_{nm}^k$ and
contortion $K_{nm}^k$
\bea
\Gamma_{nm}^k=\hat \Gamma_{nm}^k+K_{nm}^k.
\eea
The Lorentz spin connection $\gamma_{ma}^{~~b}$ and the Riemann-Cartan
connection $\Gamma_{nm}^k$ are related by the following
equation
\bea
D_m e_{an} = \pro_m e_{an} + \gamma_{ma}^{~~b} e_{bn} -\Gamma_{nm}^k e_{ak} =0.
\eea
As usually, the vielbein allows to convert
Lorentz and world indices into each other.
The torsion and curvature are defined
in a standard way
\bea
&&[D_a, D_b]=-T_{ab}^c D_c-{\bf R}_{ab},  \nn \\
&& T_{ab}^c = K_{ba}^c-K_{ab}^c,
\eea
where ${\bf R}_{ab} \equiv R_{abcd} M^{cd}$ is a Lie algebra
valued Riemann-Cartan curvature, and $M^{cd}$ is a generator of the Lorentz
Lie algebra. In component form the Riemann-Cartan curvature
$R_{mncd}$ is given by
\bea
&& R_{mncd}=\pro_n\gamma_{mcd}+\gamma_{nce} \gamma_{med} - (m\leftrightarrow n).
\eea

With these preliminaries let us
consider the concept of the gravitational vacuum
in Einstein gravity.
We will treat the vielbein $e_m^a$
as a basic field variable in Einstein gravity.
The definition of the vacuum in terms of vielbein
implies the multiple topological vacuum structure in the theory
due to the non-trivial third homotopy
group $\pi_3(SO(1,3))=\pi_3 (SO(3))=Z$
classifying non-equivalent topological
mappings $e^a_m(x): M^3 \rightarrow SO(1,3)$
\cite{cho07}. Here, we assume that the space-like hypersurface
$M^3$ has the topology of three-dimensional sphere $S^3$,
or it can be treated as $S^3$ due to compactification
of $R^3$.

The classical gravitational field
described by the metric tensor
satisfies the vacuum Einstein equation
\bea
R_{mn}-\dfrac{1}{2} R g_{mn} + \Lambda g_{mn}=0.
\eea
Due to local Lorentz invariance
the vielbein is determined by the metric,
Eqn. (\ref{def:metric}), only
up to local Lorentz transformation
$e'_{am}=L_{ab}e_{bm}$ with an arbitrary
$SO(1,3)$ matrix function $L_{ab}(x)$.

We define three types of classical
gravitational vacuum depending on
values of the cosmological constant $\Lambda $:

{\bf (I) $\Lambda=0$:} a standard notion of the
gravitational vacuum is
provided by the zero curvature condition for the Riemann tensor
\bea
&& R_{ijkl}=0.
\eea
The vacuum is defined
as a solution to the equation given
by the flat metric $g_{mn}=\eta_{mn}$.
The corresponding pure gauge vielbein
is given by an arbitrary matrix $L_{ab}(x)$
of local Lorentz transformation
\bea
\circe_{am}(x)=L_{ab}(x) \delta_{bm} = L_{am}(x). \label{pgvielbein}
\eea
Notice, since the globally defined flat metric $\eta_{mn}$
does not allow the topology of the underlying
space to be $S^3$ we don't have non-trivial topological sectors
for the corresponding vacuum vielbein. We separate the
case of possible compactification $R^3 \rightarrow S^3$
for a different definition of vacuum below.

{\bf (II) $\Lambda \neq 0$:} in the presence of the
cosmological constant the flat metric does not provide a classical vacuum solution
to Einstein equation. The vacuum can be defined by
the equations
\bea
g_{mn}=0, ~~~~~~~e_m^a=0. \label{absvac}
\eea
This vacuum represents a unique absolute vacuum in a sense that
it can be interpreted as the absence of the space-time.
This definition of the vacuum is an appropriate concept in quantum
cosmology where the space-time
can be created from "nothing" and the existence of
multiple universe is admissible as well.
The vacuum tunneling can be realized by
well-known Fubini-Study gravitational instanton
\cite{fubini} with Euler and signature numbers $\chi=3, \tau=1$
\bea
&& g_{mn}=\dfrac{4 a^2}{a^2+x^2} (\delta_{mn}-\dfrac{x_m x_n
 +\tilde x_m \tilde x_n}{a^2+x^2}), \nn \\
 && \tilde x_m = C_{mn} x^n,     \nn \\
&& x^2=\delta_{mn} x^m x^n,  \label{F-S}
 \eea
 where $C_{mn}$ is the Kaehler structure matrix and
the parameter $a$ is related to the cosmological constant
by relation $\Lambda=\dfrac{3}{2a^2}$.
The Fubini-Study metric describes
the compact space $CP^2$ without boundary.
The solution has a property: when $x^2\rightarrow \infty $
($t\rightarrow \pm \infty$) the metric vanishes, $g_{mn}\rightarrow 0$.
So that the Fubini-Study instanton describes the
vacuum-vacuum transition corresponding to the
creation and disappearance of the universe in
quantum cosmological models.
Notice, that the Fubini-Study "anti-instanton"
with $\tau =-1$ is defined
by the same metric Eqn. (\ref{F-S}) but with
opposite vielbein orientation.

Notice, that in Einstein gravity without cosmological term
the concept of the flat Minkowski metric
$\eta_{mn}$ describing an absolute space-time $R^{1,3}$ is not merely
satisfactory from the physical point of view.
An infinite space $R^{3}$  is hardly acceptable as a physical reality.
The notion of the absolute space-time
is not consistent with the second Mach principle (the well known first Mach principle relates
the inertia phenomenon with matter) stating that the
space itself is created by matter, i.e., without matter
the space is meaningless and should be absent.
In that sense the globally defined flat metric represents unphysical vacuum.
Due to these arguments we require that the physical vacuum metric
should describe a compact space, in a particular, in the present paper we
constrain our consideration of the vacuum space topology by three
dimensional spherical manifolds $S^3$
and $S^3/Z_2=RP^3 \simeq SO(3)$.

{\bf (III) $\Lambda \simeq 0$:}
we define a physical gravitational vacuum by the
locally flat vielbein $e_m^a=\eta_m^a$
on the spherical 3-manifold $S^3$ (or $RP^3$)
in the limit of infinite radius, $r\simeq \infty$.
Such a limit corresponds to infinitesimal
cosmological constant $\Lambda$. One should notice, that
this definition is not mathematically strict,
but it can serve as an adequate notion
in description of real physical phenomenona.
Such a vacuum appears in physical problems when
the space $R^3$ is compactified to $S^3$ by identifying all points at infinity
due to appropriate asymptotic boundary conditions
\cite{hawking}.

Notice, that the three dimensional sphere $S^3$ and
the projective space $RP^3$
have special features which are not available for spheres of dimension
$d \neq 3$. Namely, the spaces
$S^3$ and $RP^3$ allow the existence of almost flat
non-Riemannian connections \cite{agaoka}, for instance
at presence of contortion. With such topology of the base
space the homotopy $\pi_3(SO(1,3))$ provides
non-trivial topological vacuums
in Riemann-Cartan generalizations of  gravity.

Let us consider a simple ansatz for finding instanton
solutions. In Euclidean space-time the Lorentz group
$SO(1,3)$ is replaced by the compact group $SO(4)$.
The general pure gauge vielbein $\circe_{am}$ can be obtained from the Euclidean
flat vielbein $\delta_{am}$ by making
arbitrary Lorentz gauge transformation.
In local coordinate frame one has the same expression,
Eqn. (\ref{pgvielbein}),
for the gauge transformed vielbein as in the global
case.
Using the definition for the Levi-Civita connection (\ref{LeviCivita})
one can obtain  the corresponding pure gauge spin connection
\bea
\circphi_{mcd}=L_{ce}\pro_m \tilde L_{ed},
\eea
where $\tilde L_{ed}$ is a transposed matrix.
The Riemann tensor constructed from
the pure gauge connection is identically zero,
$\circR_{abcd}=0$.
In a temporal gauge, $\circphi_{0cd}=0$,
the static non-equivalent topological vacuums are classified by
the Chern-Simons number (winding number)
\bea
&&N_{CS}=\dfrac{1}{16\pi^2}Tr\int d^3x({\boldsymbol\circphi} d {\boldsymbol\circphi}
 +\dfrac{1}{3} {\boldsymbol\circphi}{\boldsymbol\circphi}{\boldsymbol\circphi}),
\eea
where, ${\boldsymbol\circphi}=dx^m \phi_{mcd}M^{cd}$ is a Lie algebra valued
differential 1-form of spin connection.

Let us consider a simple ansatz for
instanton configurations.
Since in Euclidean space-time the Lorentz group $SO(4)$
is locally isomorphic to the direct product
$SU(2)\times SU'(2)$ one can find a proper
generalization of the known $SU(2)$ instanton "hedgehog" ansatz.
In $SU(2)$ theory the complex scalar doublet can be parameterized
with $SU(2)$ matrix in exponential form
\bea
&&\phi = e^{i \omega \hat \tau \hat x} \phi_0, \nn \\
&&\hat x^i= \dfrac{x^i}{r},~~~~
\tan \omega = \dfrac{r}{t}, ~~~~~r=(x^i x^i)^{1/2},
\eea
where $\hat \tau^i $ are Pauli matrices,
and $\phi_0 = (0,1)$ is a trivial vacuum for $SU(2)$
scalar field.
One can write down the
following expression for a pure gauge vielbein
obtained from the trivial flat vielbein by
$SU(2)$ transformation
\bea
\circe_{ma} &=& e^{\omega \eta^i \hat x^i} \delta_{ma}=
(\delta_{ma}\cos \omega +\eta_{ma}^i \hat x \sin \omega) \nn \\
&& \cos \omega = t/ \rho, \,\,\,\, \sin \omega = r/\rho, \nn \\
&& \rho^2=t^2+r^2,
\eea
where we use 't Hooft matrices $\eta_{ma}^i, \bar \eta_{ma}^i$ (i=1,2,3).
A pure gauge vielbein constructed by the Lorentz gauge
transformation  $SO(4)\simeq SU(2)\times SU'(2)$ reads
\bea
\circe_{ma} = e^{\omega \eta^i \hat x^i} e^{\omega \bar\eta^i \hat x^i} \delta_{ma}.
\eea

In the following we will consider only one subgroup $SU(2)$
of the Euclidean Lorentz group for simplicity.
The pure gauge vielbein can be rewritten
as follows ($n=0,1,2,3$)
\bea
&&\circe_{ma} =(\delta_{ma}\cos \omega +\eta_{ma}^i \hat x \sin \omega)\equiv
\Theta_{ma}^n \dfrac{x^n}{\rho}, \nn \\
&&\Theta_{ma}^0 = \delta_{ma}, \nn \\
&&\Theta_{ma}^i = \eta_{ma}^i, \,\,\,\, i=1,2,3  \label{Theta}
\eea
where $\Theta_{ma}^n$ is a four-dimensional generalization of the 't Hooft matrices.

As a simple application of the above construction of a pure gauge
vielbein one finds a non-flat vielbein by using
a spherically symmetric "hedgehog" ansatz
\bea
e_{ma} = g(\rho) \Theta_{ma}^n \dfrac{x^n}{\rho}.
\eea
The vielbein produces a conformally flat metric $g_{mn}$
which leads to a vanishing conformal Weyl tensor
\bea
C_{mncd}&=&R_{mncd}-\dfrac{1}{2}R_{cm}g_{nd}+\dfrac{1}{2}R_{cn}g_{md}
+\dfrac{1}{2}R_{dm}g_{nc} \nn \\
&-&\dfrac{1}{2}R_{dn}g_{mc}+
\dfrac{1}{6}R(g_{cm}g_{nd}-g_{cn}g_{md})=0. \label{weyl}
\eea
The Ricci scalar
is expressed in terms of the function $g(\rho)$
\bea
&&R=2 g \big (g''+\dfrac{3g'}{\rho}\big).
\eea
For the vanishing Ricci scalar, $R=0$,  one has a simple differential
equation which has a solution
\bea
g(\rho)= 1+\dfrac{\lambda^2}{\rho^2} .\label{solconf}
\eea
This solution corresponds to the Hawking wormhole \cite{hawk2, culetu}.
The corresponding Ricci tensor is not vanished
\bea
R_{nd} = (\delta_{nd} \rho^2-4 x_n x_d)
\dfrac{4\lambda^2}{\rho^2 (\rho^2+\lambda^2)^2}.
\eea
Despite on the
seemed singularity at $\rho=0$ one can verify by using the conformal metric
with the conformal factor (\ref{solconf}) that the curvature tensor invariants
$R_{nd} R^{mn}, R_{mncd}R^{mncd}$ are regular everywhere. Since the conformal
tensor is zero, $C_{mncd}=0$, the Hirzebruch signature is zero.
That means that the solution can be interpreted as a gravitational analog to the
instanton-anti-instanton solution in Yang-Mills theory. An additional argument for such interpretation will be
given in Section IV.

{\bf III. Vacuum tunneling}

In this section we study the possibility
of tunneling between gravitational topologically
non-equivalent vacuums. For this purpose
we will introduce a construction
of gauge non-equivalent classes of vielbein different from
the one considered in the previous section.
Namely, we will choose a left-invariant basis of one-forms
on $SO(3)\simeq RP^3$ for the space triple of vielbein
expressed in terms of $SU(2)$ pure gauge connection.
The explicit construction of the instanton solution in terms
of $SU(2)$ connection allows to show explicitly that
one has vacuum tunneling.

Let us start with an explicit construction
of topologically non-trivial pure gauge connections.
The Lie algebra valued Lorentz gauge connection can be decomposed
into the 3-dimensional rotation and boost parts
$\A_m$ and $\B_m$ \cite{cho07}
\bea
{\boldsymbol \gamma}_m= \left( \begin{array}{c} \A_m \\
\B_m \end{array} \right).
\eea
Since the rotational subgroup of the Lorentz group
is locally isomorphic to $SU(2)$ one can
construct the vacuum gauge connection from the pure
gauge $SU(2)$ potential $\vOm_m$
\bea
{\boldsymbol\gamma}_m=\vOm_m=\left( \begin{array}{c} \hat \Omega_m \\
0 \end{array} \right).
\eea
Notice, that the gravitational connection of
the vacuum space-time in Einstein's theory is
fixed by the rotational part of the spin connection which describes
the multiple vacua of $SU(2)$ gauge theory \cite{baal,plb06}.

Let $\hat n_i~(i=1,2,3)$ be orthonormal isotriplets
which form a right-handed basis $(\hn_1 \times \hn_2=\hn_3)$,
and let
\bea
D_m \hn_i =0,
\label{vcon}
\eea
where $D_m$ is $SU(2)$ covariant derivative.
Obviously, these conditions impose
a strong restriction on the gauge potential
and corresponding field strength. Indeed, the constraints
(\ref{vcon}) imply a vanishing field strength.
This is because we have the following integrability condition
\bea
[D_m,~D_n]~\hn_i=g \F_{mn} \times \hn_i=0,
\eea
which leads to zero curvature equation for
$SU(2)$ field strength, $\F_{mn}=0$ ($g$ is a coupling constant).
This tells that a
vacuum potential must be the one
which parallelizes the local orthonormal frame.

Solving (\ref{vcon}) we obtain a most general $SU(2)$
vacuum potential
\bea
&\A_m=\hat \Omega_m = - C_m \hn - \dfrac{1}{g} \hn \times \pro_m \hn
= - C_m^k~\hn_k, \nn\\
&\dfrac{1}{g} \hn \times \pro_m \hn
= C_m^1~\hn_1 + C_m^2~\hn_2, \nn\\
&C_m^k = -\dfrac{1}{2g} \epsilon_{ij}^{~~k} (\hn_i \cdot \pro_m \hn_j),
\label{vac}
\eea
where $\hn=\hn_3$ and $C_m=C_m^3$.
One can easily check that $\hat \Omega_m$ describes a vacuum
\bea
&\hat \Omega_{mn} = \pro_m \hat \Omega_n
-\pro_n \hat \Omega_m + g \hat \Omega_m \times \hat \Omega_n \nn\\
&=-(\pro_m C_n^k -\pro_n C_m^k
+ g \epsilon_{ij}^{~~k} C_m^i C_n^j)~\hn_k = 0.
\label{vacf}
\eea
This tells that $\hat \Omega_m$ (or $C_m^k$)
describes the classical $SU(2)$ vacuum.
Notice that, although the vacuum is fixed by three isometries,
it is essentially fixed by $\hn$. This is because $\hn_1$
and $\hn_2$ are uniquely determined by
$\hn$, up to a $U(1)$ gauge transformation which leaves $\hn$
invariant. In general $\hat n$ describes the
Hopf fibering $\pi_3((SU(2)/U(1))=\pi_3(S^2)=Z$. We choose a special angle
parameterization for $\hat n$
\bea
&\hn = \Bigg(\begin{matrix}\sin{\alpha}\cos{\beta} \cr
\sin{\alpha}\sin{\beta} \cr \cos{\alpha}
\end{matrix} \Bigg),
\label{n}
\eea
we have the following expressions for the pure gauge vector fields $C_m^i$
\bea
&C_m^1= \oneg (\sin \gamma \pro_m \alpha
-\sin{\alpha} \cos \gamma \pro_m \beta), \nn\\
&C_m^2 = \oneg (\cos \gamma \pro_m \alpha
+\sin{\alpha} \sin \gamma \pro_m \beta), \nn\\
&C_m^3 = \oneg (\cos{\alpha} \pro_m \beta+\pro_m \gamma),
\eea
where we introduce the angle $\gamma$ corresponding to
$U(1)$ transformation which leaves $\hn$ invariant.

A nice feature of (\ref{vac}) is that the topological
character of the vacuum is naturally inscribed in it.
The topological vacuum quantum number is given by
the non-Abelian Chern-Simon index of
the potential $\hat \Omega_m$ \cite{thooft,jackiw,callan,plb79,plb06}
$(\alpha,\beta,\gamma=1,2,3)$
\bea
&N_{CS}=-\dfrac{3g^2}{8\pi^2} \int \epsilon_{\abc}
(C_\alpha^i \pro_\beta C_\gamma^i
+\dfrac{g}{3}\epsilon_{ijk} C_\alpha^i C_\beta^j C_\gamma^k) d^3x \nn\\
&=-\dfrac{g^3}{96\pi^2} \int \epsilon_{\abc} \epsilon_{ijk}
C_\alpha^i C_\beta^j C_\gamma^k d^3x, \label{nacsi}
\eea
which classifies the non-trivial topological classes.
Notice, this topology can also be described in terms of $\hn$,
because (with $\hn(\infty)=(0,0,1)$) it defines the mapping $\pi_3(S^2)$
which can be transformed to $\pi_3(S^3)$ through
the Hopf fibering \cite{plb79,plb06}. So both
$\hat \Omega_m$ and $\hn$
describe the vacuum topology of the $SU(2)$ gauge theory.
But since $\hat \Omega_m$ is essentially fixed by $\hn$ we can
conclude that the vacuum topology is imprinted in $\hn$.

Using the pure gauge $SU(2)$ vector fields $C_m^i$
one can construct the basis triple of left-invariant differential 1-forms
on $S^3$
\bea
\sigma^i=\dfrac{1}{2} dx^m C_m^i.
\eea
One can check that the one-forms $\sigma^i$ satisfy the structure Maurer-Cartan equation
\bea
d\sigma^i=2 \epsilon^{ijk} \sigma^j \sigma^k.
\eea
The basis of pure gauge vielbein one-forms can be defined in polar coordinate
system $(\rho, \theta, \phi, \psi)$ as follows
\bea
\circe {}^a=(d\rho, \rho \sigma^i).
\eea
The angle variables $\theta, \phi,\psi$ on the sphere $S^3$
have ranges
\bea
0\le \theta \le \pi, \nn \\
0\le \phi\le 2\pi, \nn \\
 0\le \psi \le 4 \pi.
 \eea
The angle functions $\alpha(\theta, \phi, \psi),\beta(\theta, \phi, \psi),\gamma(\theta, \phi, \psi)$
define the homotopy group $\pi_3(SU(2))$.
To find non-trivial
instanton solutions
one can apply the following ansatz with four
trial functions $g_0(\rho), g_i(\rho)$
\bea
e^a=(g_0(\rho) d\rho, ~~\dfrac{1}{2}dx^m g_i(\rho) \rho C_m^i).\label{ansatz}
\eea
If the functions $g_0, g_i$ are smooth then they will
provide smooth deformation of the mapping $M^4$ to $R\times S^3$.

To demonstrate the presence of quantum tunneling between non-trivial topological
vacuums we will follow
the same way as it has been done in Yang-Mills-Higgs theory \cite{plb79}.
First, one should pass to a temporal gauge. An explicit calculation
gives the following expression for the temporal
component of the pure gauge potential in Cartesian coordinates
\bea
C_t^i=\dfrac{2 x^i}{\rho^2}.
\eea
The expression for the Lie algebra valued $SU(2)$ gauge potential corresponding to the
ansatz (\ref{ansatz}) is given by
\bea
\vec A_t&=&i \dfrac{g_i(\rho)}{\rho^2} (\hat\tau^i x^i)=i\dfrac{g_i(\rho)}{\rho}(\hat\tau^i \hat x^i)\sin\omega, \nn \\
r^2&=& \sum_{i=1,2,3} (x^i)^2.
\eea
Performing gauge transformation with gauge parameters $\tilde \omega, \hat f^i$
one can impose the temporal gauge
\bea
&&\vec A_t \rightarrow \tilde A_t=\tilde U \vec A_t \tilde U^{-1}+\tilde U \vec \pro_t \tilde U^{-1}=0,\nn \\
&& \tilde U=\exp [i\tilde \omega(r,t) \tau^i \hat f^i(r,t)],
\eea
where $\hat f^2=1$.
The temporal gauge condition implies the following
equations for the gauge parameters
$(\tilde \omega, \hat f^i)$
\bea
&& \dfrac{1}{\rho} [g_i x^i \cos 2 \tomega+\sum_{j,k} \epsilon^{ijk} g_j x^j \hat f^k \sin 2 \tomega + \nn \\
&&     \sum_k  \hat f^i g_k x^k \hat f^k (1-\cos 2 \tomega) ]
-\hat f^i \pro_t \tomega-\dfrac{1}{2} \pro_t \hat f^i \sin 2 \tomega+ \nn \\
&&         \sum_{j,k} \dfrac{1}{2}\epsilon^{ijk} \hat f^j \pro_t \hat f^k (1-\cos 2 \tomega)=0. \label{temporal}
\eea
Multiplying the equation by $\hat f^i$ one finds an ordinary
differential equation for $\tilde \omega$
\bea
\pro_t \tilde \omega &=&\dfrac{1}{\rho^2} g_i x^i \hat f^i. \label{eqomega}
\eea

As an application of the above equations in temporal gauge
we consider first a simple case of the flat space-time metric
when $g_0=1$ and all functions $g_i$ are the same, $g_i=g(\rho)$:
\bea
&&ds^2=d\rho^2+\dfrac{1}{4} \rho^2 g^2(\rho) \sigma^i \sigma^i.
\eea
The zero curvature condition $R_{ijkl}=0$ implies
\bea
&& g(\rho)=1+\dfrac{2 m}{\rho}.
\eea
One can easily find a solution to the equation (\ref{temporal})
\bea
&& \hat f^i=\dfrac{x^i}{r}, \nn \\
&& \tomega=\dfrac{2mt}{r\rho}(\dfrac{m}{\rho}+2)+(1+\dfrac{2m^2}{r^2})
   \arctan \dfrac{t}{r} + c(r), \nn \\
   && c(r)=\dfrac{4m}{r}+\dfrac{\pi}{2} (1+\dfrac{2m^2}{r^2}).
\eea
   In the limit $t\rightarrow \pm \infty$ one has
\bea
   && \tomega (t=-\infty)=0, \nn \\
   && \tomega(t=+\infty)=\dfrac{8m}{r}+\pi (1+\dfrac{2m^2}{r^2}).
\eea

   This implies that $\n (t=-\infty)=(0,0,1)$ defines a trivial topology,
whereas $\n_{t=+\infty}$ corresponds to the nontrivial
topological configuration with the winding number $N_{CS}=1$
\bea
\n_{t=+\infty}=-\tilde U_{t=+\infty}
 \n_{t=-\infty} =
\left( \begin{array}{c}
\sin \alpha(r) \cos \beta(r)\\
\sin \alpha(r) \sin \beta(r)  \\
 \cos \alpha(r) \end{array} \right ),
\eea
where the angle functions $\alpha(r), \beta(r), \gamma(r)$
are related with $\tomega(r,t)$ by the equation
\bea
\tilde U_{t=+\infty}&=&\exp [i\alpha(r)/2 \hat \tau^i \hat \beta^i(r)]=\nn \\
           &&\exp [i\tilde \omega(r,+\infty) \tau^i \hat f^i(r,+\infty)], \nn \\
\hat \beta^i&=&(\sin \beta, -\cos \beta, 0).
\eea

Since the Riemann tensor is identically zero for a pure gauge connection
the total action is determined only by the surface term in the Lagrangian
of Einstein gravity
\bea
S=-\dfrac{1}{16 \pi G} \int_{M} R \sqrt {-g} d^4 x
-\dfrac{1}{8\pi}\int_{\pro M} K^i_{~i} d\Sigma,
\eea
where, $K^i_i$ is the trace of the second fundamental form
which is defined by
\bea
&&\theta^0_i=-K^i_{~j} e^i, \nn \\
&& \theta^a_{~b}=\omega^a_{~b}-(\omega_0)^a_{~b}.
\eea
The spin connection differential 1-form $(\omega_0)^a_{~b}$
is defined in the trivial flat background space-time.
Calculation of the action results in
\bea
S=3\pi m \rho_\Sigma (1-\dfrac{2m}{\rho_\Sigma})^2,
\eea
where $\rho_\Sigma$ is the radius of the
boundary surface $\pro M$ chosen as $S^3$.
The action positivness implies
$m>0$, so that the action becomes infinite in the
limit $\rho_\Sigma \rightarrow + \infty$.
By this, the transition amplitude from the trivial topological vacuum
labeled by $N_{CS}=0$
to the non-trivial one with $N_{CS}=1$ is vanished
\bea
<N_{CS}=1|N_{CS}=0>_{vac} \simeq e^{-S}=0.
\eea
Notice, in the case of Eguchi-Hanson instanton
the total action vanishes since the surface term
is proportional to $1/\rho^2_\Sigma$.

Let us consider the Eguchi-Hanson instanton solution.
The original form of the solution is the following \cite{E-H}
\bea
ds^2&=&g_0^2 d\rho^2+
 \dfrac{\rho^2}{4} (g_1^2\sigma_x^2+g_2^2\sigma_y^2+g_3^2 \sigma_z^2), \nn \\
 g_{1,2}&=&1, ~~~~~~g_3^2=\dfrac{1}{g_0^2}=1-\dfrac{a^4}{\rho^4}.
\eea
The solution has a singularity at $\rho=a$. It has been shown \cite{E-H}
that by changing the coordinate frame the solution becomes
regular everywhere for $\rho \ge a$ and in the reduced angle
range for $\psi$: $0\le \psi<2 \pi$. The point $\rho=a$
represents a removable polar coordinate singularity, and
the space-time has the topology of $RP^3$ at $\rho \rightarrow \infty$.

Notice, that the equations (\ref{temporal}) can be solved analytically
in a special case $g_1=g_2\equiv p, g_3\equiv q$ and with constrained
gauge functions $\hat f^i$ given in the form
\bea
\hat f^1&=&h \hat x, \nn \\
\hat f^2&=& h \hat y, \nn \\
\hat f^3 &=& f \hat z. \label{ansatz:f}
\eea
With this the equations (\ref{temporal}) are reduced to
\bea
\pro_t \tilde \omega &=& \dfrac{r q}{\rho^2 f}, \nn \\
\pro_t f &=& 0, \nn \\
f&=& \dfrac{r}{{\sqrt{z^2+\dfrac{p^2}{q^2}(x^2+y^2)}}}~. \label{analytic}
\eea
The solution implies an additional
constraint for the trial functions $p,q$
\bea
\pro_t (\dfrac{ p^2}{q^2})=0.
\eea
To find a solution to the equations (\ref{temporal}) in the case of
Eguchi-Hanson instanton one has to introduce three independent gauge functions
$\hat f^{1,2}, \tilde \omega$. So that one has to modify the ansatz (\ref{ansatz:f})
which will be given below. Notice,
in the asymptotic region $t\rightarrow \pm \infty$
the function $g_3^2=1-\dfrac{a^4}{\rho^4}$ goes to the flat limit very fast,
so that one can use the analytic solution (\ref{analytic}) for qualitative
analysis of the asymptotic behavior of $\hat f^i, \tilde \omega$
\bea
f&\simeq&1, \nn \\
\tilde \omega &\simeq& \int dt \dfrac{rg_3}{\rho^2}+c_1(r).
\eea
The function $c_1(r)$ has to be chosen from
the initial condition $\tilde \omega(t=-\infty)=0$.
At upper limit one has
\bea
\tilde \omega (t=+\infty) \simeq \pi + \omega_0 (r),
\eea
where $\omega_0 (r\rightarrow \infty)=0$. So that, the Eguchi-Hanson
instanton realizes the tunneling from the trivial
vacuum with $N_{CS}=0$ to the non-trivial one with $N_{CS}=1$.

Let us consider a consistent parametrization for three independent gauge functions
$\hat f^{1,2}, \tilde \omega$
which implies a solution to the equations (\ref{temporal})
for the case of Eguchi-Hanson instanton with manifest axial symmetry.
The proper parametrization is obtained by generalization of the special ansatz (\ref{ansatz:f})
by rotating the functions $\hat f^{1,2}$ in the plane $x^1,x^2$ and introducing
time dependence for the gauge functions
\bea
\hat f^1&=&P(x,t) x^1+Q(x,t) x^2, \nn \\
\hat f^2&=&-Q(x,t) x^1+P(x,t) x^2.
\eea
The third independent gauge function $\tilde \omega$
remains the same. Substitution of this ansatz into
equations (\ref{temporal},\ref{eqomega}) results in the
following equations ($g_{0,3}=1/g_{1,2}\equiv g(\rho)$)
\bea
&&\dfrac{\cos 2 \tomega}{g \rho^2}+\dfrac{g}{\rho^2} x^3 \sin 2\tomega Q
-P \cos 2 \tomega \pro_t \tomega \nn \\
  &&              -\dfrac{1}{2} \sin 2 \tomega \pro_t P+\dfrac{1-\cos 2 \tomega}{2}
(\hat f^3 \pro_t Q-Q \pro_t \hat f^3)=0, \nn \\
&& \dfrac{\sin 2 \tomega}{g \rho^2}(\hat f^3 -g^2 x^3 P) -Q \cos 2 \tomega \pro_t \tomega -
\dfrac{1}{2} \sin 2 \tomega \pro_t Q \nn \\
&& +\dfrac{1-\cos 2 \tomega}{2} (P\pro_t \hat f^3-\hat f^3 \pro_t P)=0, \nn \\
&& \dfrac{g}{\rho^2} x^3 \cos 2 \tomega - \dfrac{\sin 2 \tomega}{g\rho^2} u^2 Q
-\cos 2 \tomega \hat f^3 \pro_t \tomega \nn \\
&&-\dfrac{1}{2} \sin 2 \tomega \hat f^3-\dfrac{1-\cos   2 \tomega}{2}u^2 (\pro_t PQ-P\pro_tQ) =0, \nn \\
&& \pro_t \tomega=\dfrac{1}{\rho^2} \Big (\dfrac{1}{g}u^2(P^2+Q^2)^{1/2}+gx^3 \hat f^3\Big ), \label{eqfs}
\eea
where $u^2\equiv (x^1)^2+(x^2)^2$. Notice that there are only three independent
equations in (\ref{eqfs}). The equations contain coefficients which depend
only on cylindric coordinates $x^3, u$, this implies axial symmetry of the
solutions $P,Q,\tomega$ under rotation around the axis $x^3$.

There is another useful parametrization for the gauge functions
$\hat f^{1,2}, \tomega$
which is suitable for numerical solving the equations (\ref{temporal}, \ref{eqomega}).
It is convenient to choose angle parametrization
for the functions
\bea
\hat f^k (t,\vec r)=
\left( \begin{array}{c}
\sin \alpha(t,\vec r) \cos \beta(t,\vec r)\\
\sin \alpha(t,\vec r) \sin \beta(t,\vec r)  \\
 \cos \alpha(t,\vec r) \end{array} \right ).
\eea
Numerical testing of the equations (\ref{temporal},\ref{eqomega}) confirms
regular behavior of the functions $\alpha(t,\vec r), \beta(t,\vec r), \tilde \omega(t,\vec r)$
in the whole region $\rho \ge a,~~(-\infty<t<\infty)$.
The solution for the gauge functions is depicted in Fig. 1.
For simplicity we show the regular behavior of the gauge functions
along the selected radial direction $x^1=x^2=x^3=r/\sqrt 3$ (for positive values of $x^i$).
We have checked numerically that the gauge function $\tomega(t,r)$ which
implies the non-trivial winding number has a correct limit $\tomega(t\rightarrow +\infty)=\pi$.
This completes our proof that Eguchi-Hanson instanton
provides tunneling between non-trivial topological neighbor vacuums
with the base space topology of $RP^3$.

\begin{figure}[t]
\includegraphics[scale=0.5]{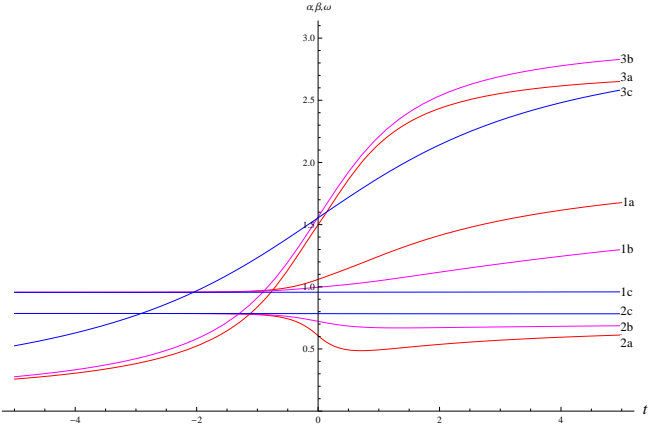}
 \caption{\label{Fig. 1} The curves (1a,b,c) and (2a,b,c) correspond to the gauge functions
$\alpha(t,r)$ and $\beta(t,r)$ respectively. The curves (3a,b,c) depict the behavior of the
gauge function $\tomega(t,r)$ which has correct asymptotic limits $(0,\pi)$. The size parameter of
Eguchi-Hanson instanton is set to be $a=1$. The subscripts (a,b,c) correspond to the fixed values of
the space radius: $r=1, 1.2, 3$. The initial values for the gauge functions are taken to be close
to the following asymptotic values $\alpha(t=-\infty)=\pi/4,~~\beta(t=-\infty)=\arccos (x^1/r),
~~\tomega(t=-\infty)=0$.}
\end{figure}

{\bf IV. Self-dual contortion}

In generalized gravity models with contortion (torsion)
the total Riemann-Cartan curvature can be decomposed
into two parts in accordance with the split relationship (\ref{split})
for the spin connection
\bea
&& R_{abcd}=\hat R_{abcd}+\tilde R_{abcd}, \nn \\
&& \hat R_{abcd}=\hat D_{\underline b} \varphi_{\underline a \underline c
\underline d}+\varphi_{bc}^{~~\,\,e}\varphi_{aed}-(a\leftrightarrow b), \nn \\
&& \tilde R_{abcd}=\hat D_{\underline b} K_{\underline a \underline c\underline d}
+K_{bc}^{~~e} K_{aed}-(a\leftrightarrow b) , \label{decomp2}
\eea
where, $\hat R_{abcd}$ is the Riemann curvature and $\hat D_a$
is a restricted covariant derivative containing only the Levi-Civita connection.
The underlined indices stand for indices over which the
covariantization is performed.
Due to curvature  decomposition (\ref{decomp2})
the classical vacuum can be defined by several ways.
A simple definition of the vacuum in generalized Riemann-Cartan gravity
includes two zero curvature conditions
\bea
\hat R_{abcd}&=&0, \nn \\
\tilde R_{abcd}&=&0.
\eea
So that, in the space-time with a flat
metric the tunneling is possible
due to instanton configurations made of contortion.
Non-trivial topological classes of contortion are provided by the
same homotopy group $\pi_3(SO(1,3))$ as in Einstein gravity with vielbein.
In this Section we consider possible configurations of self-dual contortion
irrespectively on a concrete model of generalized Riemann-Cartan gravity.
For simplicity,
we suppose the vielbein to be flat, $e_m^a=\eta_m^a$. So that,
$\gamma_{mcd}=K_{mcd}$ and the
total Riemann-Cartan curvature $R_{abcd}$ coincides with the
curvature $\tilde R_{abcd}$.

For the Riemann-Cartan curvature one can define
two types of dual tensors using contraction
of the antisymmetric tensor $\epsilon_{abcd}$
with either first or second index pair of $R_{mncd}$
\bea
R^*_{mnab}&=&\dfrac{1}{2} \epsilon_{abcd} R_{mncd}, \nn \\
{}^*R_{mncd}&=&\dfrac{1}{2} \epsilon_{mnkl} R_{klcd}.
\eea
We define a self-dual Riemann-Cartan curvature
as a tensor satisfying the double self-duality equations
\bea
R_{mncd}&=&R^*_{mncd}, \nn \\
R_{mncd}&=&{}^*R_{mncd}.
\eea
Using the 't Hooft matrix $\eta^i_{cd}$ one can decompose
any antisymmetric tensor $T_{ab}$ into self-dual and anti-self-dual
parts
\bea
T_{ab}=\eta^i_{ab} S^i_{ab} + \bar \eta^i A^i_{ab}.
\eea
A self-dual Riemann-Cartan curvature can be written
in the following form
\bea
R_{mncd} = \eta_{cd}^i R^i_{mn}.
\eea
The solution to the self-duality condition is
provided by the self-dual spin connection
with arbitrary functions $\gamma^i_{m}$
\bea
\gamma_{mcd} = \eta^i_{cd}\gamma^i_{m}.
\eea
In the case of Riemann geometry the self-duality
condition implies the following expression for the Riemann curvature
\bea
R_{mncd}&=&\eta^i_{mn} \eta^j_{cd} P^{ij}, \nn \\
\eta_{mn}^i &\equiv& e^a_m e^b_n \eta^i_{ab},\label{dualansatz}
\eea
where the tensor $P^{ij}$ must be symmetric due to
the symmetry  of the Riemann tensor under the replacement
of first and second index pairs.
The self-dual Riemann-Cartan curvature has the same
form (\ref{dualansatz}) with a non-symmetric
tensor $P^{ij}$ in general.
Let us construct some double self-dual contortion
configurations using a proper ansatz.

I. We apply the ansatz
\bea
\gamma_{mcd} = \eta^i_{cd} \bar \eta^i_{mn} x^n f(\rho).
\eea
After substituting this ansatz into the Eqn. (\ref{dualansatz})
one can find
\bea
&& \eta^i_{mn} P^{ij} = -2 \bar \eta^j_{mn} (f+\rho^2 f^2) \nn \\
&&+(\bar \eta^j_{mp} x_n-\bar \eta^j_{np} x_m) x_p (\dfrac{f'}{\rho}-2 f^2).
\eea
Self-duality condition of the equation implies the constraint
\bea
\bar \eta^k_{mn} \eta^i_{mn} P^{ij}=0.
\eea
The last equation gives an ordinary differential  equation
\bea
4f+\rho f' + 2 \rho^2 f^2=0
\eea
which has a solution
\bea
f(\rho) = - \dfrac{\lambda^2}{\rho^2 (\rho^2+\lambda^2)}.
\eea
This solution is analog to 't Hooft-Polyakov one instanton solution
in a singular gauge. Notice, that the tensor $P^{ij}$ is not symmetric
\bea
P^{ij}=- 2 \eta^i_{mn} \bar \eta^j_{mk} x^n x^k \dfrac{f}{\rho^2+\lambda^2},\label{instII}
\eea
 so that the curvature $R_{mncd}$
represents essentially the Riemann-Cartan curvature.
The contracted Riemann-Cartan curvatures are not vanished
\bea
&& R_{mncn}=\delta_{mc} P^{ii}+\eta^k_{mc} \epsilon^{kij} P^{ij}, \nn \\
&& R=4 P^{ii}.
\eea

II. We choose the following  ansatz
\bea
\gamma_{mcd} = \eta^i_{cd} \eta^i_{mn} x^n Q(\rho).
\eea
The solution to double self-duality equations for $R_{mncd}$ reads
\bea
Q=-\dfrac{1}{a^2+\rho^2}.
\eea
The curvature tensors have the following forms
\bea
&&R_{mncd} = \eta_{mn}^i \eta_{cd}^i \dfrac{2a^2}{(\rho^2+a^2)^2}, \nn\\
&& R_{ab} = \delta_{ab} \dfrac{8 a^2}{(\rho^2+a^2)^2}, \nn\\
&& R=\dfrac{32 a^2}{(\rho^2+a^2)^2}.
\eea
The solution can be interpreted as a solution to the Riemann-Cartan
analog of the Einstein equation
with a non-constant cosmological term
\bea
R_{ab}=\Lambda(\rho) \delta_{ab}.
\eea

III. Let us construct a self-dual solution to self-duality
condition for the conformal tensor $C_{mncd}$ (\ref{weyl})
defined in terms of Riemann-Cartan curvature.
We use the following ansatz
\bea
\gamma_{ncd} = \eta^i_{cd}\eta^i_{nk} x^k f(\rho)
 + \bar \eta^i_{cd}\bar \eta^i_{nk} x^k f(\rho). \label{ansIII}
\eea
Substituting the ansatz into the self-duality condition
for the conformal tensor and requiring the condition
of vanishing scalar Riemann-Cartan curvature $R=0$
one obtains a differential equation
\bea
&&R=-12(4 f(\rho)+2 \rho^2 f^2(\rho) + \rho f'(\rho))=0,
\eea
which has a solution
\bea
f(\rho) = - \dfrac{\lambda^2}{\rho^2 (\rho^2+\lambda^2)}. \label{solIII}
\eea
This solution implies
\bea
&&C_{mncd}=0, \nn \\
&&R=0, \nn \\
&& R_{nd} = (\delta_{nd}\rho^2-4 x_n x_d) \dfrac{4\lambda^2}{\rho^2 (\rho^2+\lambda^2)^2}. \label{ricci}
\eea
The solution is regular everywhere with a finite curvature invariant
\bea
R^2_{mncd} = \dfrac{1152\lambda^4}{(\rho^2+\lambda^2)^4}.
\eea

The ansatz (\ref{ansIII}) contains two parts,
each of them corresponds to self-dual contortion described
by type II solution. So that the solution (\ref{solIII})
can be interpreted as an analog to the instanton anti-instanton pair.
Notice, that this solution is very similar
to the conformally flat metric (\ref{solconf}) considered in Section II.
Notice, if we adopt the point of view that torsion
is responsible for the microscopic structure of the space-time,
and our Universe represents a classical macroscopic system, we can
perform averaging procedure in the solutions (\ref{ricci}) and (\ref{instII})
over all directions using the averaging prescription
\bea
<x_n x_m>= \dfrac{1}{4} \rho^2 \delta_{nm}.
\eea
This implies vanishing of the Riemann-Cartan curvature.
By this way the contortion may become unobservable at macroscopic level.

{\bf V. Discussion}

We have shown explicitly that the Eguchi-Hanson
instanton can provide tunneling between
non-trivial topological vacuums represented by
the vielbein in the case when the base space has topology of $RP^3$.
Our main assumption is that the vielbein
represents a more fundamental variable
than the metric tensor.
It might seem unexpected that the vacuum tunneling
requires the space topology of $RP^3$, not $S^3$. An interesting
discussion on which topology of the base space, $S^3$ or $RP^3$,
should be accepted as a physical one, is
presented in Ref. \cite{McInnes}.

One should notice, that in our present Universe
the vacuum tunneling is unlikely to be available
since the Universe is not static and represents rather a
macroscopic, noncoherent system in a quantum sense.
However there is a possibility for experimental detecting
the non-trivial vacuum structure. It is related to
the presence of Adler-Bardeen-Jackiw (ABJ) axial anomaly,
in a similar manner with quantum chromodynamics.
A non-vanishing signature leads to
the axial anomaly of the axial current
for spin $\dfrac{1}{2}$ and spin $\dfrac{3}{2}$
particles.
In the case of Eguchi-Hanson instanton
the spin index $I_{1/2}$ of the Dirac operator
is identically zero
whereas the spin index $I_{3/2}$
is non-trivial. For the case of
Fubini-Study instanton one has
an axial ABJ anomaly \cite{fubini}
\bea
\pro_\mu j^{5\mu}=\pro_\mu (\sqrt g e^{a\mu} \bar \psi \gamma_a \gamma_5 \psi)
 = \dfrac{1}{4} RR^*.
 \eea

Unfortunately in quantum gravity
we don't have an analogue of the pion
decay coupling constant $f_\pi$ which
could allow to measure the axial charge in gravity.
As for the index $I_{3/2}$, there is a hope that
spin 3/2 particles (like the $\Omega^-$ hyperon or
the hypothetic particle gravitino) could open a way
of direct detecting the non-trivial vacuum topology.
Besides these pure thoughtful speculations one should notice that
one has one indirect evidence why our space might
have the topology of $RP^3$. This may come from
the existence of the positive cosmological constant which provides the
$RP^3$ topology of space-like hypersurfaces
of the closed de Sitter universe.

The main assumption which we explore in our consideration
of vacuum tunneling in Einstein gravity
is that the vielbein represents a fundamental variable
responsible for the quantum gravitational effects.
If the Einstein gravity is an emergent phenomenon,
i.e., it is an effective theory, then one should not
quantize the vielbein which is a pure classic
object by its geometric origin.
Since a consistent quantum theory of gravity is unknown for the present moment,
the question whether the vielbein or the torsion (or another object) is responsible
for the quantum dynamics of gravity, remains open.
One possible way to testify whether the vielbein can be
more fundamental at quantum level than the metric is
to perform experiment on detecting the gravitational analog of Aharonov-Bohm effect.
We hope to study the theoretical framework for this in nearest future.
Recently we have considered a gravity model with a topological phase
where the vielbein does not play a role in quantum dynamics
whereas the contortion (as a part of Lorentz connection)
plays a fundamental role at quantum level \cite{pakijmp}.
In this connection our self-dual torsion configurations
may have some physical applications.

{\bf Acknowledgements}

One of authors (DGP) thanks Dr. Alex Nielsen for numerous useful discussions.
The work is supported in part by Korean National Research Foundation
(2008-314-C00069 and 2010-002-15640) and by the Brain Pool Program (032S-1-8) of KOSEF.

\end{document}